\documentclass[aps,pre,showpacs,twocolumn]{revtex4}
\usepackage{graphics}
\usepackage[dvips]{color}
\usepackage[T1]{fontenc}
\usepackage[latin1]{inputenc}
\usepackage{graphicx}
\usepackage{amssymb}
\usepackage{rotating}

\begin{document}
\input epsf.sty


\title{The density of states of classical
spin systems with continuous degrees of freedom}

\author{Andreas Richter}
\affiliation{Institut f\"ur Theoretische Physik I,
Universit\"at Erlangen-N\"urnberg, D -- 91058 Erlangen, Germany}
\affiliation{Max-Planck-Institut f\"ur Kolloid- und Grenzfl\"achenforschung,
D-14424 Potsdam, Germany}

\author{Michel Pleimling}
\affiliation{Institut f\"ur Theoretische Physik I,
Universit\"at Erlangen-N\"urnberg, D -- 91058 Erlangen, Germany}

\author{Alfred H\"{u}ller}
\affiliation{Institut f\"ur Theoretische Physik I,
Universit\"at Erlangen-N\"urnberg, D -- 91058 Erlangen, Germany}

\begin{abstract}
In the last years different studies have revealed the usefulness of a microcanonical
analysis of finite systems when dealing with phase transitions. In this approach
the quantities of interest are exclusively expressed as derivatives of the 
entropy $S = \ln \Omega$ where $\Omega$ is the density of states. Obviously, the
density of states has to be known with very high accuracy for this kind of analysis.
Important progress has been achieved recently in the computation of the density of
states of classical systems, as new types of algorithms have been developed. Here we
extend one of these methods, originally formulated for systems with discrete degrees
of freedom, to systems with continuous degrees of freedoms. As an application we
compute the density of states of the three-dimensional $XY$ model and
demonstrate that critical quantities can directly be determined from the density
of states of finite systems in cases where the degrees of freedom take continuous values.

\end{abstract}
\maketitle

\section{Introduction}
In the microcanonical treatment of a finite system the main quantity
of interest is the specific entropy $s(e,m)$ as a function of the
energy $e$ and of the order parameter $m$ \cite{gros01}.
The latter is given by the magnetization in the case of a magnet.
Recent investigations of finite classical systems with discrete degrees 
of freedom undergoing a phase transition in the 
infinite volume limit have revealed that
the microcanonically defined spontaneous magnetization
$m_{sp}(e)$ is zero for energies larger than a certain energy $e_c$
and rises steeply with a power law behaviour for energies
smaller than $e_c$ \cite{kast00,huel02}.
Interestingly,
the corresponding susceptibility, being directly related to the
curvature of the entropy surface, diverges at $e=e_c$. 
It has to be noted that the exponents governing the power law behaviour
of the different quantities in the vicinity of $e_c$ 
take on the classical mean field values for all
system sizes smaller than infinite \cite{behr04}. The true nonclassical
exponents of the infinite system can, however, be determined
from a microcanonical scaling analysis \cite{plei04}.
Whether such a behaviour with a sharp onset of the order parameter
and a diverging susceptibility is termed a ``phase transition in
the microcanonical ensemble'' or not, is a semantic question and,
as such, of lesser importance than the question if there really
exists a point with a true divergence already for a finite
system. This has frequently been challenged with the reasoning
that for a discrete system as e.g. the Ising model the arguments
of the entropy assume only discrete values. For an Ising system
with nearest neighbour interactions
on a $d$-dimensional hypercubic lattice with linear extension $L$ these values are
$e_k=E_k/N$ and $m_l=M_l/N$ with $E_k = 4kJ$ and $M_l = 2l$,
where $k$ and $l$ are integers, $J$ is the coupling constant
and $N=L^d$ is
the number of spins in the system. Only in the limit of infinite
system size do the ratios of differences become equal to the
derivatives (with respect to $e$ and $m$) needed for the calculation
of the susceptibility and other physical quantities.

In order to overcome these serious objections against the
microcanonical way of analyzing critical phenomena, we have
decided to determine the entropy of a classical spin system 
where the spins take on continuous values. Of
course, there the entropy $s(e,m)$
is a continuous function of its arguments. The system
chosen is the $XY$ model in $d=3$ dimensions which undergoes a second
order phase transition. 

In recent years different numerical methods have been proposed for the
computation of the density of states of classical models \cite{oliv98,wang98,wang00,wang01,huel02}.
Here we consider a highly efficient algorithm which has been 
applied to several discrete models as e.g. the two- and three-dimensional
Ising models \cite{huel02}, the three-states Potts model \cite{behr04,behr05},
the vector Potts model with four states \cite{behr04} or the voter model \cite{sast03}.
For the computation of the density of states of systems with continuous degrees
of freedoms we have to modify this algorithm.

The outline of the paper is the following. In Section 2 we generalize
the numerical method presented
in \cite{huel02} to models with continuous variables. 
This method yields the transition variables which permit the 
construction of the entropy surface. 
It is important to note that 
transition variables can also be obtained when using other algorithms
as for example the Wang-Landau method \cite{wang01}. Therefore our method for
deriving the entropy from these quantities can be applied very generally.
Data obtained in this way 
for the three-dimensional $XY$ model are analyzed
microcanonically in Section 3. This enables us to determine
critical exponents in systems with continuous variables
directly from the density of states, either by extrapolating
effective exponents or from a microcanonical finite-size scaling
ansatz. Section 4 gives our conclusions.

\section{The computation of the microcanonical entropy}

\subsection{Transition variables}

The algorithm developed in the following allows the determination of the 
density of states (DOS) $\Omega$ (and therefore also of the microcanonical entropy)
of classical spin systems.
In an extension of the method introduced in \cite{huel02},
the computation of the microcanonical entropy is performed in the case 
where the DOS is a function of continuous variables.
The method is exemplified for the three-dimensional $XY$ model
with the classical Hamiltonian
\begin{equation} \label{eq_xy}
{\mathcal H} = - \sum\limits_{\langle i,j \rangle} \, \vec{S}_i \cdot \vec{S}_j
\end{equation}
where the spin $\vec{S}_i$, characterizing the lattice point $i$ of a simple cubic lattice, 
is a two-dimensional 
vector lying on the unit circle. The sum in (\ref{eq_xy}) is over nearest
neighbour bonds.

\begin{figure}
\centerline{\epsfxsize=3.25in\ \epsfbox{
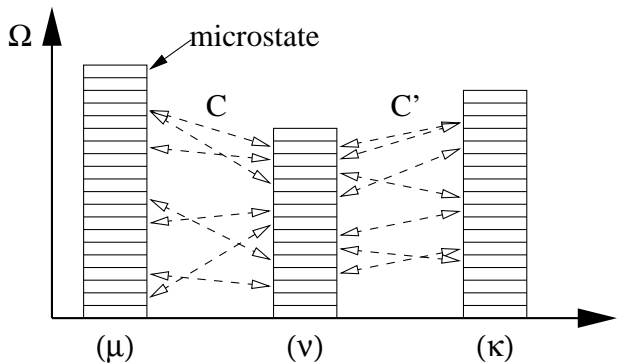}
}
\caption{\label{fig1} The density of states at different macrostates $\mu$, $\nu$ and $\kappa$ (schematic). 
For a reversible mechanism the number of connections 
$C=C(\mu \rightarrow \nu)=C(\nu \rightarrow \mu)$ between two macrostates is the same for the forward and 
backward transitions.}
\end{figure}

The consideration underlying our method is relatively simple. Assume at
first a classical discrete spin system. A macrostate of the system may for
convenience be denoted by $\mu$. One might think of the energy $e_k$ and magnetization
$m_l$ that characterize a macrostate or level $\mu = (e_k,m_l)$. In general, a huge
number of microstates belongs to a macrostate $\mu$. This number is the
degeneracy $Z_{\mu}$ of the level $\mu$. 
In the course of the simulation a
reversible mechanism takes the system from a microstate in a level $\mu$
to a different one which belongs to another level $\nu$. This step is repeated many times.
Of course, the same mechanism has to be applied at every update. Starting from one
microstate, a number $\mathcal{N}$ of new microstates can be generated. The number $\mathcal{N}$
depends on the mechanism and on the model under consideration.

When the mechanism operates on all microstates of the level $\mu$ then
$\mathcal{N}Z_{\mu}$ new microstates can be generated. A number $C(\mu \rightarrow \nu)$
of these belong to the level $\nu$. The quantity of specific interest is
$w(\mu \rightarrow \nu)=C(\mu \rightarrow \nu)/\mathcal{N}Z_{\mu}$
which is the probability of arriving at anyone of the microstates belonging
to the level $\nu$ when the starting point was one of the microstates of level $\mu$. As the
mechanism is reversible the numbers of forward and backward transitions between
the levels $\mu$ and $\nu$ are the same: $C(\mu \rightarrow \nu)=C(\nu \rightarrow \mu)$,
see Figure \ref{fig1}.
This then yields the expression
\begin{equation}
\frac{Z_\mu}{Z_\nu} = \frac{w(\nu \rightarrow \mu)}{w(\mu \rightarrow \nu )}.
\label{f_omega_1}
\end{equation}
which allows the determination of the degeneracies $Z_\mu$ from the transition probabilities $w$.
Eq.\ (\ref{f_omega_1}), which is sometimes called Broad Histogram Equation \cite{oliv98},
is already complete for discrete spin models like the Ising or Potts models.

Let us  now focus on spin models with continuous spin variables.
As an example we discuss the $XY$ model \cite{muno99}, but the generalization to other models (as for example
the Heisenberg model) is straightforward.
In our case both the energy and the magnetization have continuous values and the magnetization is a two-dimensional vector.
As the energy function (\ref{eq_xy}) is invariant under global spin rotations, it is sufficient
for many investigations to consider only the modulus $m$ of the magnetization.
This also considerably reduces the amount of resources needed for the numerical simulation of the model.
For the purpose of performing this simulation we discretize the energy and the magnetization,
the discretized values being denoted by $\hat{e}$ and $\hat{m}$, and call
$\delta e$ respectively $\delta m$ the width of the discretization.
We thereby allocate all microstates with energies $e$ between $\hat{e} - \delta e/2$ and
$\hat{e} + \delta e/2$ and modulus $m$ of the magnetizations between $\hat{m} - \delta m/2$ and
$\hat{m} + \delta m/2$ to the same macrostate denoted by $\mu = (\hat{e},\hat{m})$.
As the magnetisation ${\bf{m}}=(m_1,m_2)$ is a two-dimensional vector the ``volume'' 
\begin{equation} \label{eq_V}
V(\hat{m}) = 2\pi\, \hat{m}\, \delta e\, \delta m
\end{equation}
in ($e$,$m$)-space is not a constant. The DOS
$\Omega(\hat{e},\hat{m})$ is derived from
\begin{equation} \label{eq_dos}
\Omega(\hat{e},\hat{m}) = Z(\hat{e},\hat{m})\, / \, V(\hat{m})
\end{equation}
where $Z(\hat{e},\hat{m}) \equiv Z_{\mu}$.
This expression also holds for other models beyond the $XY$ model, where one
only has to replace the volume (\ref{eq_V}) by the appropriate expression. It is important to 
note that during the simulation the spin variables as well as the energy and the magnetization of the system adopt continuous values.
The discretisation only concerns the quantities depending on the macrostates, e.~g.~the DOS 
and the transition probabilities.

As already mentioned we perform
the simulation with a reversible mechanism.
For the $XY$ model 
we use single spin rotations of randomly selected spins and a random rotation angle $-\pi \leq \varphi <\pi$.
The following description of the algorithm is very general and also applies to discrete
spin models.
Suppose that the system is in a macrostate denoted by $\hat{e}$ and $\hat{m}$ or equivalently by $\mu$.
In the next step it is attempted to bring the system into another state $\nu$ by the use of the mechanism described above.
We increase the number of
attempts $B(\mu)$ to leave the macrostate $\mu$ by one count.
At the same time we add one count to $T(\mu \rightarrow \nu)$ which is the number of attempted 
transitions from $\mu$ to $\nu$.
In a long run the ratio $t(\mu \rightarrow \nu):=T(\mu\rightarrow \nu)/B(\mu)$ finally approximates 
the transition probability $w(\mu\rightarrow \nu)$.
It is the quantity $t(\mu\rightarrow \nu)$ which we call transition variable.
During the simulation $B(\mu)$ and $T(\mu\rightarrow \nu)$ are updated at every attempted step.
The probability of acceptance of the transition from $\mu$ to $\nu$ is chosen to be
\begin{equation}
p:=\min\left[\frac{t(\nu\rightarrow \mu)}{t(\mu\rightarrow \nu)};1\right]
\label{f_p}
\end{equation}
This choice of $p$ leads to an equal number of attempts $B(\mu)$ 
to leave any macrostate $\mu$, if sufficiently many updates of the system are performed.
This is due to the fact, that now the probabilities for an actually executed transition for both the forward and backward directions are equal.
It is worth mentioning that the probability of acceptance changes during the simulation and that 
it approaches its asymptotic value 
\begin{equation}
\tilde{p} =\min\left[\frac{w(\nu\rightarrow \mu)}{w(\mu\rightarrow \nu)};1\right]
\label{f_pasym}
\end{equation}
for very long runs.

\subsection{Construction of the entropy surface}
Suppose that a macrostate $\mu=(\hat{e},\hat{m})$ whose degeneracy $Z_\mu$ 
has not yet been determined can be
reached from several macrostates $\nu$ and suppose that the values of the $Z_\nu$ 
are known, then with the knowledge of the transition probabilities one may calculate
$\ln Z_\mu$ starting from one of the macrostates with $\nu = \nu'$:
\begin{displaymath} \label{omega}
\ln Z_\mu = \ln \frac{w(\nu' \rightarrow \mu)}{w(\mu \rightarrow \nu')} + \ln Z_{\nu'}
\end{displaymath}
and one would obtain the same result  $\ln Z_\mu$
for each of the states $\nu$ as a starting point.
However, the simulation only yields the transition variables $t(\mu \rightarrow \nu)$
which are estimates for the transition probabilities $w(\mu \rightarrow \nu)$. Consequently, as
these estimates are subject to stochastic fluctuations, we end up with different values
for $\ln Z_\mu$ when starting from different macrostates $\nu$.

To take this into account
we propose to estimate $\ln Z_\mu$ by the weighted average 
\begin{equation} \label{bh4}
\ln Z_\mu = \sum\limits_\nu p_{\mu \nu} \, \left(
\ln  \frac{t(\nu \rightarrow \mu)}{t(\mu \rightarrow \nu)}
+ \ln Z_\nu \right)
\end{equation}
where the sum is over all macrostates $\nu$ from which the macrostate $\mu$ can be reached.
The weights $p_{\mu \nu}$ are given by
\begin{equation} \label{pj}
p_{\mu \nu}= \frac{g_{\mu \nu}}{\sum\limits_\kappa g_{\mu \kappa}}
\end{equation}
with
\begin{equation} \label{g2}
g_{\mu \nu} = \frac{ T(\mu \rightarrow \nu) \,  T(\nu \rightarrow \mu) }{T(\mu \rightarrow \nu) +
T(\nu \rightarrow \mu) }
\end{equation}
which follows from the gaussian error propagation.
Another possible choice for the $g_{\mu \nu}$ is given by
\begin{equation} \label{g1}
g_{\mu \nu} = \min \left( T(\mu \rightarrow \nu); T(\nu \rightarrow \mu) \right).
\end{equation}
We checked that both choices (\ref{g2}) and (\ref{g1}) lead to microcanonical entropies which,
within the errors, are the same.

Up to now we have tacitly assumed that the degeneracies $Z_\nu$ entering the sum in Eq.\
(\ref{bh4}) were known exactly. This is of course not the case, the $Z_\nu$ being also affected
by statistical errors. To reduce the errors in the entropy values we propose the following
iterative scheme. At the start we attribute an arbitrary
value $\ln Z_{\nu_0}=C_0$ 
to a randomly chosen macrostate $\nu_0$.
Note that the choice of $C_0$
does of course not affect our analysis as microcanonically defined physical quantities
only involve derivatives of the microcanonical entropy \cite{gros01}. Furthermore, 
if one would use the generated entropy in a canonical analysis, the constant $C_0$
would drop out when taking canonical averages of any observable. 
Having chosen an initial macrostate, we perform a random walk in the parameter
space (in our case the energy-magnetization space) and attribute to any visited macrostate $\mu$
the degeneracy $Z_\mu$ obtained by evaluating Eq.\ (\ref{bh4}). Of course, in doing so,
only macrostates which have in the past been visited at least once can contribute 
to the sum in Eq.\ (\ref{bh4}).
This is repeated until every macrostate has often been visited, typically a few
thousand times. When $\ln Z_\mu$ has been obtained for all the macrostates $\mu = (\hat{e},
\hat{m})$, Eq.\ (\ref{eq_dos}) yields the
microcanonical entropy as (setting $k_B=1$)
\begin{equation} \label{smicro}
S(\mu)=\ln Z_\mu - \ln V(\hat{m}).
\end{equation}

\begin{figure} 
\centerline{\epsfxsize=3.25in\ \epsfbox{
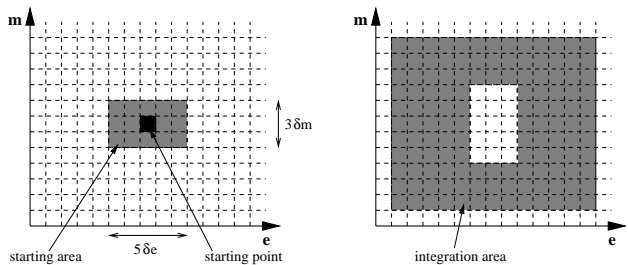}
}
\caption{\label{fig2} The integration area at the beginning (left figure) and after several 
extensions (right figure). For this sketch we supposed that the initial point is only reachabel
from microstates inside the grey area (left). At later stages, the middle of the integration
area is not changed anymore (right).}
\end{figure}

It is of advantage not to integrate directly over the whole parameter space, but
to restrict the random walk at the beginning to a small area around the starting point $\nu_0$, as shown in Figure \ref{fig2}.
The area at the start must be large enough to encompass all the macrostates reachable
from the macrostate $\nu_0$.
Typically after a few thousands
visits of every macrostate inside the initial area, this integration area is
increased. This change of integration area is performed periodically.
At some stage we no longer update the center of the integration area, 
see Figure \ref{fig2}.

The presented method has the advantage that {\it all} transition variables computed
in the simulation are used for the construction of the entropy. Furthermore,
the iterative scheme just described ensures that the error in the final result
for the entropy is minimized. It must be noted that this method is of course
also applicable for other models, including models with discrete spins,
as only the correct volume $V$ has to be inserted into Eq.\ (\ref{smicro}).
Finally, we mention that for larger system sizes it is often of advantage
to compute the entropy of stripes restricted in energy direction \cite{huel02,schu03}.
The adaption of the presented method to this
restricted geometry in the parameter space is straightforward.

\section{Analysis of the entropy}
\begin{figure} 
\centerline{\epsfxsize=3.25in\ \epsfbox{
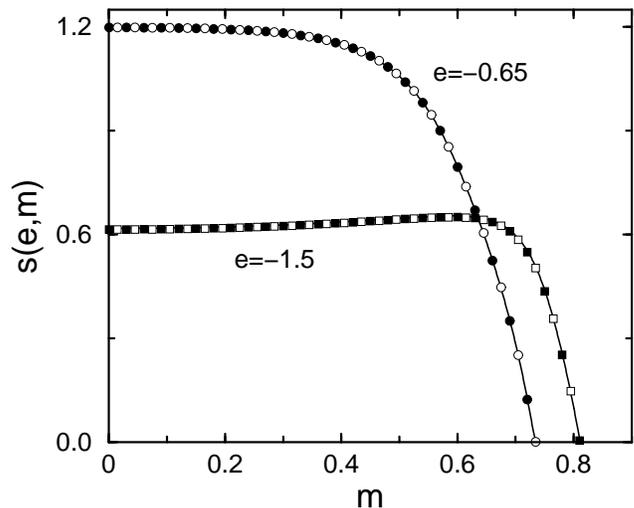}
}
\caption{\label{fig3}
Computed entropy for the three-dimensional $XY$ model with $N=10^3$ spins 
as function of the modulus of the magnetization for two different
energies and two different discretizations: $\delta e = \delta m = 1/N$ (full symbols) and
$\delta e = \delta m = 2/N$ (open symbols). The data have been obtained by simulating two different stripes
centered at the energies $e=-0.65$ and $e=-1.5$. Only selected points are shown by symbols.
Note that for $e=-0.65$ the entropy is maximal for $m=0$, whereas for $e=-1.5$ the maximum is located at
$m = m_{sp,L}(e) > 0$. For illustrative reasons
the data have been shifted vertically
by adding a constant. 
}
\end{figure}

The procedure outlined in the previous Section has been applied
to systems of different linear extension $L$ up to $L$ = 25.
For most of the runs the size of the channels has been
chosen to be unity on the extensive scale, i.e. $\delta e = \delta m = 1/N$,
though different discretizations have also been tried with
no noticeable effect on the results. For $L=10$ this is illustrated in Figure \ref{fig3}
where we show the computed entropy as a function of $m$ for two different 
fixed values of the energy and two different discretizations.
The data obtained for the different discretizations perfectly agree within errors.
Note that this is an important point as the independence of the results on
the bin size demonstrates that we are indeed accessing continuous properties.

Whereas for small systems the entropy $s(e,m)$ can be computed in a single run,
for larger systems the increasing number of channels in energy and in magnetization
direction makes the computation of the entropy very time consuming.
One can then restrict the determination of the transition
variables to narrow stripes restricted in the energy direction, which permits a
distribution of the computation on different CPUs.
The width of our stripes
was usually 25 units on the extensive energy scale, which corresponds
to twice the maximum energy increment in a single move.
On the intensive scale the stripes are rather narrow
$(\approx 3 \cdot 10^{-3})$ for a system with $L=20$.
Therefore the resulting entropies
can be averaged over the 25 energy channels in order to
obtain $s(e, m)$ at the center $e$ of the stripe.

Having the entropy or, equivalently, the density of states 
at our disposal we can use these data in different ways.
Of course, we can compute the canonical partition function from the density of states
and then obtain thermal averages of the different quantities of interest as
the energy, the susceptibility and so on. Here, we take a different route and directly
investigate the microcanonical entropy $s(e,m)$ itself. Recent investigations 
of discrete spin systems \cite{huel02,plei04,behr05,hove04} have shown that critical exponents
governing the power law behaviour of quantities in the vicinity of a critical point
can be determined reliably by analyzing directly the density of states of finite systems.
Here we extend this analysis to phase transitions taking place in classical
spin systems with continuous degrees of freedom \cite{hove04b}. This kind of approach also poses a
severe test to our method of computing the entropy, as a microcanonical analysis
requires data of extremely high quality due to the absence of
the smoothening effect of the Boltzmann weights.

\begin{figure} 
\centerline{\epsfxsize=3.25in\ \epsfbox{
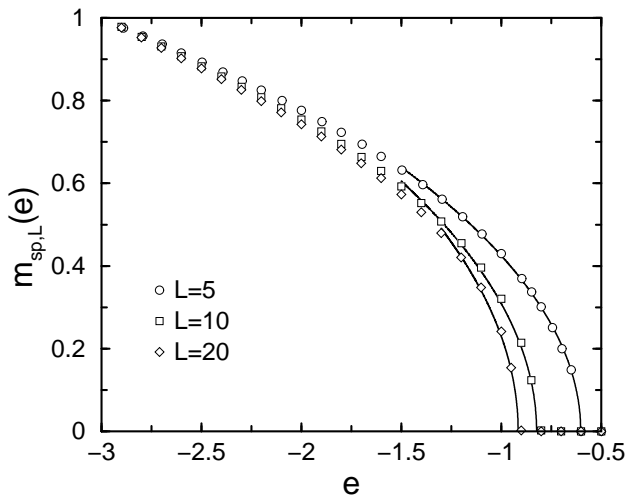}
}
\caption{\label{fig4}
Spontaneous magnetization of the three-dimensional $XY$ model
as function of the energy per spin for three different sizes. Note the
sharp onset of the spontaneous magnetization at the size-dependent
energy $e_{c,L}$. Error bars are much smaller than the sizes of the symbols.
Lines are obtained by fitting the data close to $e_{c,L}$ to 
a square root (\ref{eq:msp}).
}
\end{figure}

Coming back to our numerical data we observe that the entropy $s(e,m)$,
as a function of the modulus of the magnetization, shows one
maximum at $m = 0$ for $e \geq e_{c,L}$ and one maximum at
$m = m_{sp, L}(e) > 0$ for $e < e_{c,L}$, see Figure \ref{fig3}.
For a given energy $e$ the DOS as function of $m$ 
exhibits an extremely sharp maximum such
that the overwhelming majority of the accessible states
belongs to the value of $m$ where $s$ has a maximum.
Therefore $m_{sp,L}(e)$ is identified with the
spontaneous magnetization of the finite system \cite{kast00,huel02,behr04,plei04}.
Figure \ref{fig4} shows the spontaneous magnetization
for three system sizes. In order to produce this plot
the maximum has been determined for a series of stripes
centered at different values of $e$.
Notice the sharp onset of $m_{sp,L}(e)$ at $e_{c,L}$,
a behaviour also observed for discrete spin models displaying
a continuous phase transition in the infinite volume limit \cite{kast00,huel02,plei04,behr04}.
This onset is accompanied by a divergence of the radial part of the microcanonically defined susceptibility \cite{behr04b}
\begin{equation} \label{eq:chi}
\chi(e,m)=- \frac{\frac{\partial s}{\partial e} \frac{\partial^2 s}{\partial e^2}}{\frac{
\partial^2 s}{\partial e^2} \frac{\partial^2 s}{\partial m^2} - \left( 
\frac{\partial^2 s}{\partial e \partial m} \right)^2 }.
\end{equation}
This divergence is due to the fact that the curvature 
in the radial direction of $m$ (i.e. the denominator in (\ref{eq:chi}))
vanishes at $e_{c,L}$.

\begin{figure} 
\centerline{\epsfxsize=3.25in\ \epsfbox{
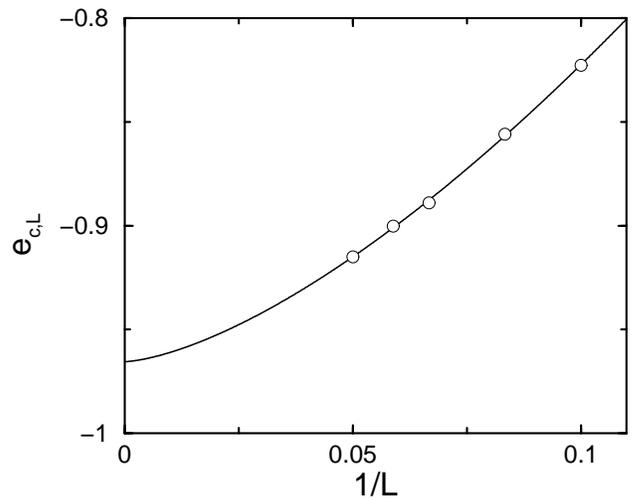}
}
\caption{\label{fig5}
Pseudo-critical energy $e_{c,L}$, defined by the sharp 
onset of the spontaneous magnetization, as function of the inverse linear extension $1/L$.
The full line results from a fit of Eq.\ (\ref{eq:ecl}) to the data points, with
$e_{c,\infty} = -0.965$, $B=4.534$ and $\nu_\epsilon= 0.666$. Error bars are comparable to the
symbol size.
}
\end{figure}

Exploiting the fact that close to $e_{c,L}$ the
spontaneous magnetization rises with a square root
singularity, $e_{c,L}$ can be determined with high
precision from a fit of the nonzero values of
$m_{sp,L}(e)$ by the expression
\begin{eqnarray} \label{eq:msp}
m_{sp,L}(e)= A_L (e_{c,L}-e)^{1/2}
\end{eqnarray}
where $A_L$ and $e_{c,L}$ are adjustable parameters. The resulting
values of $e_{c,L}$ are plotted in Figure \ref{fig5}
as a function of $1/L$ together
with the fit function
\begin{eqnarray}  \label{eq:ecl}
e_{c,L} = e_{c,\infty}+ B(1/L)^{\nu_\epsilon}
\end{eqnarray}
with the adjustable parameters $e_{c,\infty}$, $B$ and $\nu_\epsilon$. The fit yields
the values $e_{c,\infty} = -0.965(10)$
and $\nu_\epsilon= 0.666(5)$ which,
considering the relatively small systems
we have simulated, are in good agreement with the best values
found in the literature, namely $e_c = -0.9884$ \cite{cucc02}
and $\nu = 0.67155(27)$ \cite{camp01}.

We briefly pause to recall that in the entropy formalism considered here critical
exponents are not always identical to the better known thermal critical exponents.
Indeed, in cases where the specific heat is diverging at the critical point (implying that
the thermal critical exponent $\alpha$ governing this divergence is positive,
$\alpha > 0$) critical exponents $x_\epsilon$ appearing in the microcanonical analysis
are related to their canonical counterparts $x$ by $x_\epsilon = x/(1 - \alpha)$ \cite{kast00}. This is 
not so in cases where the specific heat displays a cusp singularity with $\alpha < 0$.
There $x_\epsilon = x$. For the three-dimensional $XY$ model we have $\alpha \approx - 0.015$ \cite{camp01}
and consequently critical exponents obtained in the microcanonical analysis are identical
to the thermal critical exponents. We can therefore savely drop the index $\epsilon$
from now on.

\begin{figure} 
\centerline{\epsfxsize=3.25in\ \epsfbox{
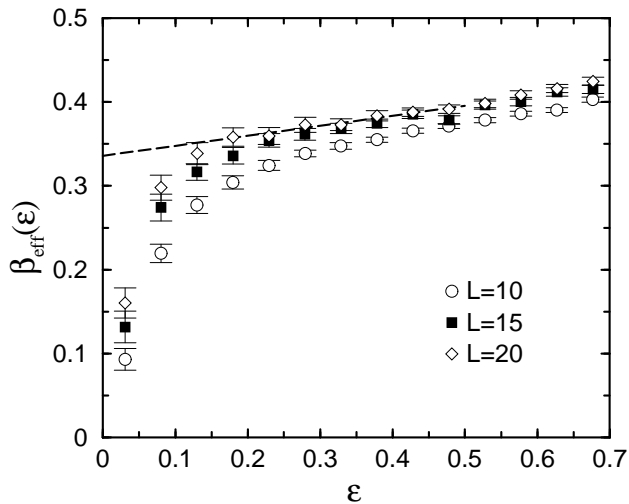}
}
\caption{\label{fig6}
Effective exponents derived from the spontaneous magnetization, see Eq.\ (\ref{eq:beff}),
as function of $\varepsilon$ for different system sizes. The dashed line 
extrapolates the data for $L=20$ with $0.15 < \varepsilon < 0.50$ to zero.
}
\end{figure}

\begin{figure} 
\centerline{\epsfxsize=3.25in\ \epsfbox{
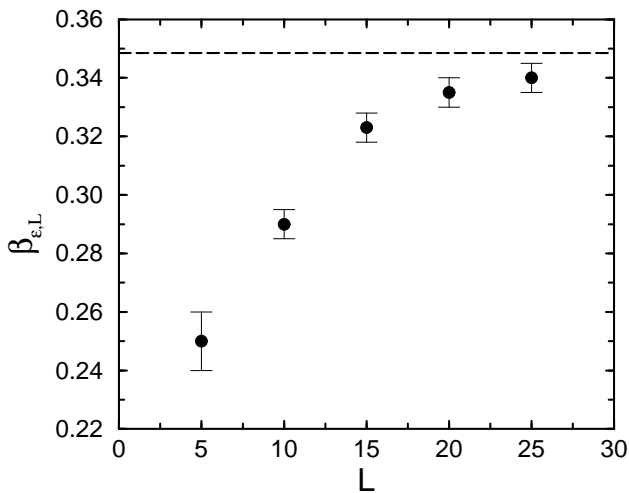}
}
\caption{\label{fig7}
Variation of $\beta_{\varepsilon,L}$ obtained from extrapolating linearly the effective 
exponents $\beta_{eff}$ as function of $L$. The dashed line indicates the literature
value $\beta=0.3485$ \cite{camp01}. 
}
\end{figure}

For all finite system sizes the expansion (\ref{eq:msp}),
which is valid in the vicinity of $e_{c,L}$,
yields the classical critical exponent
$\beta_{c,L} = 1/2$ \cite{behr04}, but the range of validity
of (\ref{eq:msp}) shrinks to zero with $L \rightarrow \infty$.
Outside of this range $m_{sp,L}(e)$ differs very little from
$m_{sp}(e) = m_{sp,\infty}(e)$ for not too small system sizes.
The critical expansion
for the infinite system is given by
\begin{eqnarray} \label{mspcrit}
m_{sp}(e) = A\, \varepsilon^{\beta}
\end{eqnarray}
where $\varepsilon = \frac{e_c-e}{e_c-e_g}$ and $e_g = -3$
is the ground state energy per degree of freedom. 
Eq.\ (\ref{mspcrit}) is stricly speaking only valid
in the limit $\varepsilon \longrightarrow 0$.
One may analyze the spontaneous magnetization
by looking at the energy dependent
logarithmic derivative \cite{huel02}
\begin{eqnarray} \label{eq:beff}
\beta_{eff}(\varepsilon) =
\frac{d \ln m_{sp}(\varepsilon)}{d \ln(\varepsilon)}
\end{eqnarray}
which approaches the true critical exponent
$\beta$ in the limit
limit $\varepsilon \rightarrow 0$. Plots which show this
approach for the two- and three-dimensional Ising systems can
be found in \cite{huel02}.
Figure \ref{fig6} shows
$\beta_{eff}(\varepsilon)$ of the $XY$ model under study
for several lattice sizes. All the graphs, of course,
precipitate to zero on approaching $\varepsilon = 0$
because $m_{sp,L}(e)$ is finite at $e = e_c$, i.e.
for $\varepsilon = 0$, but the region of constant slope
around $\varepsilon = 0.5$ can be used to extrapolate the data to $\varepsilon = 0$.
Figure \ref{fig7} shows the result of this extrapolation
for five values of L.
These values quickly come close to the expected value
$\beta=0.3485(2)$ \cite{camp01}.

\begin{figure}
\centerline{\epsfxsize=3.25in\ \epsfbox{
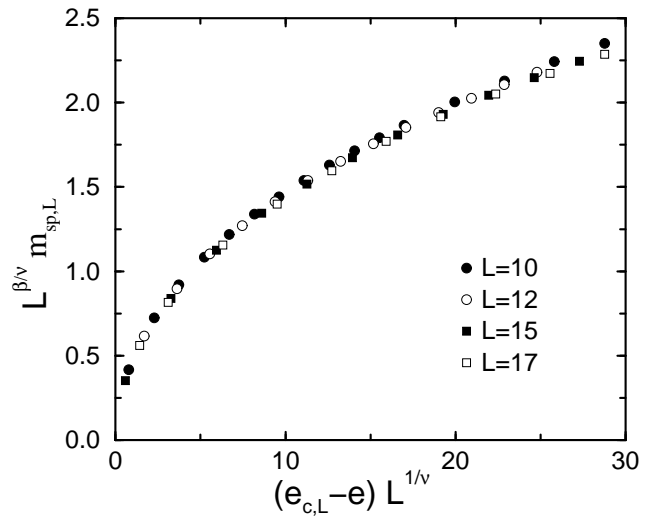}
}
\caption{\label{fig8}
Microcanonical finite-size scaling plot for the three-dimensional $XY$ model.
The values $\beta/\nu=0.53(1)$ and $1/\nu=1.47(2)$ result from the best 
data collapse. Error bars are comparable to the sizes of the symbols.
}
\end{figure}

Microcanonical finite-size scaling, as suggested in \cite{plei04}, is another way
of determining critical exponents. This approach takes advantage of the
existence of a well-defined transition point $e_{c,L}$ in finite microcanonical
systems and leads to the scaling ansatz
\begin{equation} \label{gl3}
L^{\beta/\nu} \, m_{sp,L}\left( ( e_{c,L} -e )
\right) \sim  W \left( C \, ( e_{c,L} -e ) \, L^{1/\nu}
\right)
\end{equation}
where $W$ is a universal scaling function characterizing the given universality class
and $C$ is a nonuniversal constant.
It is worth noting that in this approach the knowledge of the value of the critical energy $e_c$
of the infinite system is not needed. As shown in \cite{plei04} Eq. (\ref{gl3}) yields
reliable values of the involved critical exponents for discrete models. We show in Fig.\ \ref{fig8}
that this is also the case for the continuous model discussed here. Plotting 
L$^{\beta/\nu} \, m_{sp,L}$ against $( e_{c,L} -e ) \, L^{1/\nu}$, we obtain 
the values  $\beta/\nu=0.53(1)$ and $1/\nu=1.47(2)$ 
from the best data collapse, in good agreement with
the expected values \cite{camp01} 0.5189 and 1.4891.

\section{Conclusions}
In recent years impressive progress has been achieved in the numerical
computation of the DOS of finite classical systems, as new and highly 
efficient simulation methods have been proposed. One of the aims of this
work is the generalization of the method presented in \cite{huel02}
to systems with continuous degrees of freedom. This method relies
on the powerful concept of transition variables \cite{oliv98}. In 
our implementation the actual values of the transition variables are
used for the acceptance rates of a proposed move, and this during
the entire run. As a consequence all sites of the parameter space
are visited with the same rate.

With the final values of the transition variables we can construct
the density of states, using the complete information contained in
these variables. We have formulated this approach in the present paper
for magnetic systems, where the natural variables are given by energy
and magnetization, but a generalization to other situations 
is straightforward.

With the knowledge of the density of states we can analyze the system under
investigation in different ways. One common approach is to compute the
partition function and then proceed with a canonical analysis. The good quality
of our data, however, also enables us to perform a microcanonical analysis where
derivatives of the entropy with respect to energy and magnetization 
play a predominant role. The usefulness of the microcanonical analysis
in the study of phase transitions has been revealed in many recent studies
\cite{gros01,huel02,plei04,gros00,huel94,behr05}.

The theory of phase transitions is usually
formulated within the frame of the canonical ensemble,
which is based on the partition function or on its logarithm,
the free energy. All the thermodynamic functions, as e.g. the
susceptibility, are calculated from derivatives of the free
energy with respect to the temperature and the applied fields.
For all finite system sizes these derivatives remain finite,
but in the thermodynamic limit they may diverge at
the critical point, if a continuous phase transition occurs
in the system.

In the microcanonical ensemble the analysis of phase
transitions is based on the DOS or its logarithm, the entropy \cite{gros01,behr04,gros00}.
It has been demonstrated for several discrete spin models
that the abrupt onset of the order parameter at $e_c$ and the
divergence of the susceptibility occurs already for systems
consisting of a rather modest number of spins, albeit with
the classical values of the critical exponents \cite{kast00,behr04}. 

We have demonstrated in this work that similar features are observed 
in systems with continuous spins:
In a finite system of linear extension $L$ the curvature of the entropy
surface in the magnetization direction changes sign
at the point $e = e_{c,L}$. It is there where the order
parameter sets in and where the susceptibility diverges
in continuous as well as in discrete systems. 
Therefore for the $XY$ model, where $e$ and $m$ take continuous values,
the entropy $s(e,m)$ shows the same behaviour as for discrete spin models
where it is defined only at discrete points in the parameter space.

Of course, numerical studies must be performed at discrete
values of the variables. The distinguishing fact is that 
for continuous models these values
can be chosen freely, such that derivatives can, in principle,
be calculated from the ratios of arbitrarily small differences.
The reality of a numerical study, however, sets the limits.

As a final application we have demonstrated that in systems with continuous 
degrees of freedom critical exponents can in principle
be determined directly from the density of states, along the same lines
as in discrete models \cite{huel02,plei04,behr05,hove04}. In particular, we
obtain very good estimates of $\nu$ and $\beta$ with regard to the modest system sizes
used in this study.

\acknowledgments
We thank the Regionales Rechenzentrum Erlangen for
the use of the IA32 compute cluster. We also thank Hans Behringer for
useful discussions and comments.


\begin{thebibliography}{99}
\bibitem{muno99} The reader should note that in \cite{muno99b} an implementation
of the Broad Histogram method to systems with continuous degrees of freedom
has been presented. This implementation
differs in many aspects from our algorithm. For example, Mu\~{n}oz and Herrmann
did not use the actual values of the transition variables for the acceptance
or the rejection of a proposed spin rotation. Furthermore, they
did not try to compute the density of states as a function of both the energy and
the magnetization.
\end{thebibliography}
\end{document}